# The record low thermal conductivity of monolayer Cuprous Iodide (CuI) with direct wide bandgap

Jinyuan Xu[1], Ailing Chen[1], Linfeng Yu[1], Donghai Wei[1], Qikun Tian[2], Huimin Wang[3],

Zhenzhen Qin[2], and Guangzhao Qin[1,*]

[1]*State Key Laboratory of Advanced Design and Manufacturing for Vehicle Body, College of Mechanical and Vehicle Engineering, Hunan University, Changsha 410082, P. R. China*

[2]*International Laboratory for Quantum Functional Materials of Henan School of Physics and Microelectronics, Zhengzhou University, Zhengzhou 450001, P. R. China*

[3]*Hunan Key Laboratory for Micro-Nano Energy Materials & Device and School of Physics and Optoelectronics, Xiangtan University, Xiangtan 411105, Hunan, P. R. China*

**Abstract:** Two-dimensional materials have attracted lots of research interests due to the fantastic properties that are unique to the bulk counterparts. In this paper, from the *state-of-the-art* first-principles, we predicted the stable structure of monolayer counterpart of the $\gamma$-CuI (Cuprous Iodide), which is a *p*-type wide bandgap semiconductor. The monolayer CuI presents multifunctional superiority in terms of electronic, optical, and thermal transport properties. Specifically, the ultralow thermal conductivity of 0.116 Wm$^{-1}$K$^{-1}$ is predicted for monolayer CuI, which is much lower than $\gamma$-CuI (0.997 Wm$^{-1}$K$^{-1}$) and other typical semiconductors. Moreover, an ultrawide direct bandgap of 3.57 eV is found in monolayer CuI, which is larger than $\gamma$-CuI (2.95-3.1 eV), promoting the applications in nano-/optoelectronics with better optical performance. The ultralow thermal conductivity and direct wide bandgap of monolayer CuI as reported in this study would promise its potential applications in transparent and wearable electronics.

---

* Correspondence author <gzqin@hnu.edu.cn>

# 1. Introduction

Since the discovery of graphene[1], two-dimensional (2D) materials have attracted lots of research interests due to the fantastic properties. In particular, materials in the 2D form usually possess better performance than the bulk counterparts, and the properties regulation in 2D materials can be more effective due to the larger surface volume ratio. Up to now, numerous 2D materials with outstanding properties have been reported and investigated, such as silicene[2,3], phosphorene[4], MoS$_2$[5], *etc.*, which have shown potential applications in electronics[6], optoelectronics,[7] catalysis[8], thermoelectrics[9], *etc.*

In the microelectronics revolution, the semiconductors with wide bandgap take key position. For instance, the Gallium Nitride (GaN) has been widely used for high-power electronics[10] and the blue light LEDs[11], which was awarded the 2014 Nobel in Physics[12]. In addition, the Zinc Oxide (ZnO)[13] is also a widely used *n*-type semiconductor in the field of transparent electronics with the direct wide bandgap of 3.4 eV. Among the candidates for the potential applications in transparent electronics, the *n*-type semiconductors usually have small effective mass while *p*-type semiconductors usually have large effective mass. However, it is found that the bulk Copper Iodide in cubic wurtzite (γ-CuI) is a *p*-type semiconductor with small effective mass, which promises the high carrier mobility[14] and would benefit the applications coupled with *n*-type semiconductors. For instance, the γ-CuI is found to possess potential applications in thermoelectrics[15] due to the large Seebeck coefficient.

Considering the successful cases in the past of the extraordinary properties of 2D materials as compared to the bulk counterparts, it is expected that the monolayer CuI could have better properties than the γ-CuI. As a non-layered I-VII group compound, CuI crystallizes into three different phases: α, β, and γ. The temperature changes can cause the phase transition of CuI, *i.e.*, from the cubic γ-phase to the hexagonal β-phase with temperature above 643 K, and the β-phase can be further transformed to the cubic α-phase with temperature above 673 K[16]. Thus, the CuI based structures are abundant depending on different conditions. In fact, the ultrathin 2D nanosheets of γ-CuI has already been synthesized very recently [17]. However, it is in the film form of the non-layered γ-CuI, and the structure of the single-layer CuI would be different from the unit-layer in the γ-CuI film due to the dimension limit. Thus, the structure and the stability are necessary to be comprehensively investigated for monolayer CuI. Moreover, the promising physical properties of monolayer CuI is expected to promote the applications and advances in optoelectronics and thermoelectric fields.

In this paper, from the *state-of-the-art* first-principles, we predicted the stable structure of monolayer CuI and presented the multifunctional superiority by performing systematical study on the electronic, optical, and thermal properties. A direct bandgap is found in monolayer CuI, which is larger than the γ-CuI, leading to the ultra-high optical transmission. Moreover, ultralow thermal conductivity is predicted for monolayer CuI, which is 1~2 orders of magnitude lower than most semiconductors. The direct wide bandgap and the ultralow thermal conductivity of monolayer CuI would promise its potential applications in transparent and wearable electronics.

## 2. Results and discussion

### 2.1 The structure and the stability of monolayer CuI

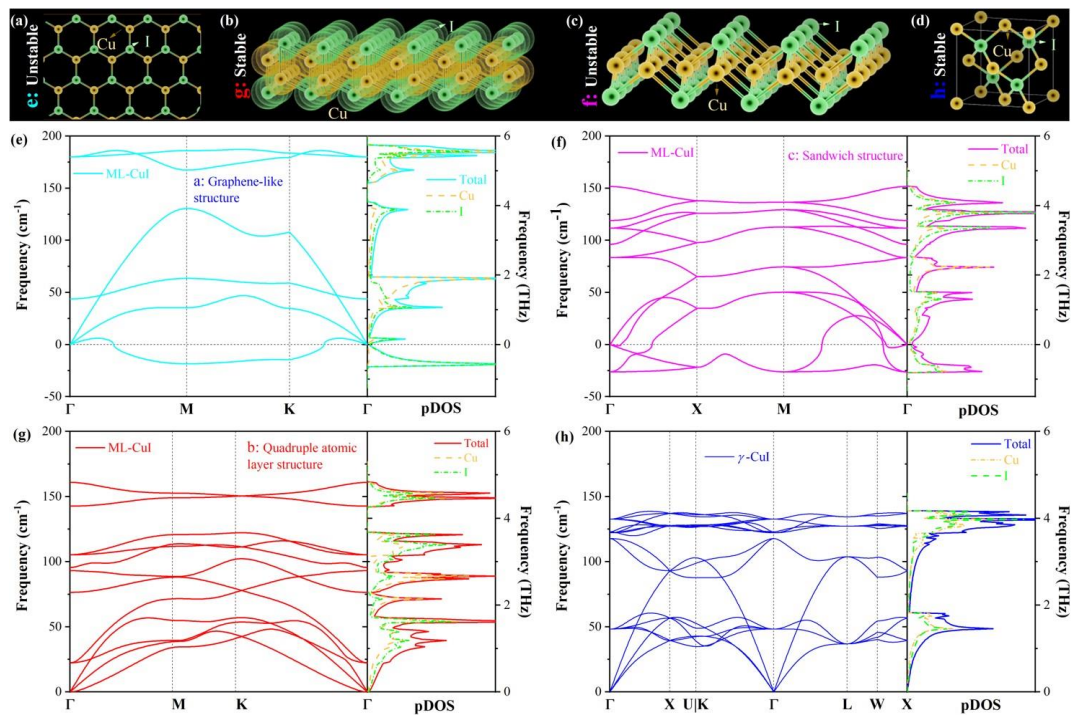

Figure 1. The stable structure of monolayer CuI is verified by the phonon dispersions. The possible structures of monolayer CuI (marked as ML-CuI): (a) graphene-like structure, (b) quadruple atomic layer structure, (c) sandwich structure. (d) The stable $\gamma$-phase bulk structure (marked as $\gamma$-CuI). (e-h) The phonon dispersion curves correspond to the structures shown in (a-d). The partial density of states ($p$DOS) is also presented.

As reported in previous studies, 2D materials are commonly formed in a few typical structures[18], such as the hexagonal honeycomb structure represented by graphene, the hinge-like puckered structure represented by phosphorene, the sandwich structure represented by $MoS_2$, the buckled structure represented by silicene, *etc*. Thus, for exploring the probably stable structure of monolayer CuI (ML-CuI), instead of the global searching strategies[19,20], it would be of high efficiency to construct the possible monolayer CuI in the pattern of the common structures in 2D, and further verify the structural stability based on the lattice dynamics by calculating the phonon dispersions.

By testing all the possible patterns of structures for 2D materials, it is found that negative frequency exists for monolayer CuI except for the buckled structure as shown in Fig. 1(c). The monolayer CuI in the planar hexagonal honeycomb structures, similar to graphene and sandwich structure, are presented here in Fig. 1(a,b) as examples for contrast, where the negative frequencies in the phonon dispersion reveal the unstable nature [Fig. 1(e,f)]. Thus, by examining the stability of monolayer CuI in different patterns of 2D structures, the stable structure of monolayer CuI is successfully found with two buckled sublayers, showing similar characteristics to silicene. The optimized monolayer CuI crystallizes in the hexagonal structure with space group of $P\bar{3}m1$, which is the minimum energy state among all the possible structural states of monolayer CuI. The optimized lattice constants are $a = b$ = 4.18 Å for monolayer CuI, which is in good agreement with experimental result (4.19 Å) as reported by Mustonen, K. *et al*[21]. In contrast, the bulk $\gamma$-CuI crystallizes in zinc blende structure with space group of $F\bar{4}3m$, and the optimized lattice constants $a = b = c = 6.08$ Å as obtained in this work is in good agreement with the results (5.99-6.03 Å) in literatures[22, 17].

There are four atoms in the primitive cell of monolayer CuI and the vibration frequency is low, which may lie in the heavy atomic masses. By examining the $p$DOS of the lattice vibration, it is found that the Cu and I atoms contribute near equally at different frequencies. Moreover, there exists a bandgap among the optical phonon branches [Fig.1(c)], which may lead to the weakened scattering of optical phonon modes as reported previously. In contrast, there is no phonon frequency bandgap in the $\gamma$-CuI [Fig.1(d)].

### 2.2 The ultralow thermal conductivity

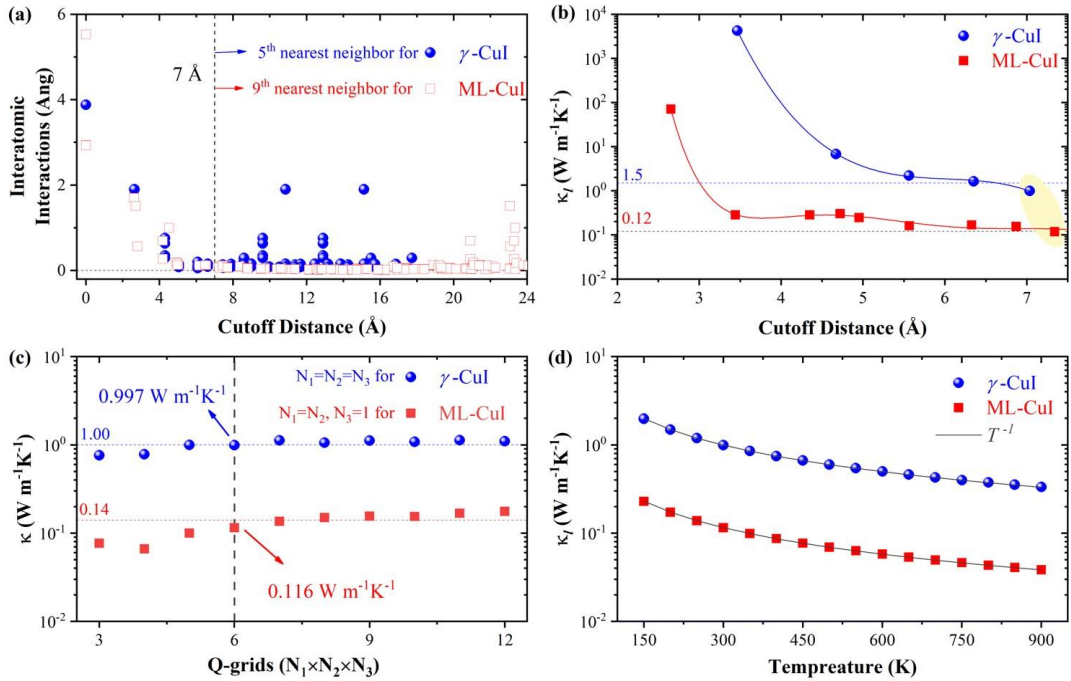

Figure. 2. The thermal conductivity and the convergence tests of related parameters. (a) The variation of interatomic interactions with respect to the atomic distance. (b) The convergence of thermal conductivity with respect to the cutoff distance. (c) The convergence of thermal conductivity with respect to the $Q$-grids. (d) The thermal conductivity of monolayer CuI and $\gamma$-CuI as a function of the temperature.

The thermal transport properties of monolayer CuI and $\gamma$-CuI are comparably investigated based on the stable structures, which is crucial for its practical applications in devices. The lattice thermal conductivity ($\kappa_l$) is calculated by solving the phonon Boltzmann transport equation (BTE) based on first-principles calculations with no adjustable parameters. It is well known that the main heat carriers in semiconductors are phonons and the contribution of the electronic thermal conductivity is negligible. Thus, the thermal conductivity is discussed in terms of lattice thermal conductivity in the following.

The convergence of thermal conductivity is verified based on the analysis of interatomic interactions[23] [Fig. 2(a)]. As shown in Fig. 2(b), the thermal conductivity decreases with the increasing cutoff distance, where a few stages are presented. The decreasing thermal conductivity is due to the inclusion of more interatomic interactions and more phonon-phonon scattering. Note that the thermal conductivity still presents a decrease with cutoff distance larger than 6 Å, which reveals the significant effect of long-range interactions in monolayer CuI. Such long-range interactions usually exist in materials with resonant bondings, such as phosphorene[24] and PbTe[25]. With a full convergence test, the thermal conductivity of monolayer CuI is predicted as 0.116 W $m^{-1}K^{-1}$ [Fig. 2(c)] at 300 K, which is a record low value close to the thermal conductivity of air. The ultralow thermal conductivity of monolayer CuI is much lower than the most of already known semiconductors. The calculated thermal conductivity of $\gamma$-CuI is 0.997 W $m^{-1}K^{-1}$, which is in excellent agreement with the ~0.55 W $m^{-1}K^{-1}$ reported by Yang $et$ $al$[15]. Noted that the temperature-dependence of thermal conductivity for both monolayer CuI and $\gamma$-CuI are perfectly matching the $1/T$ decreasing relation [Fig. 2(d)].

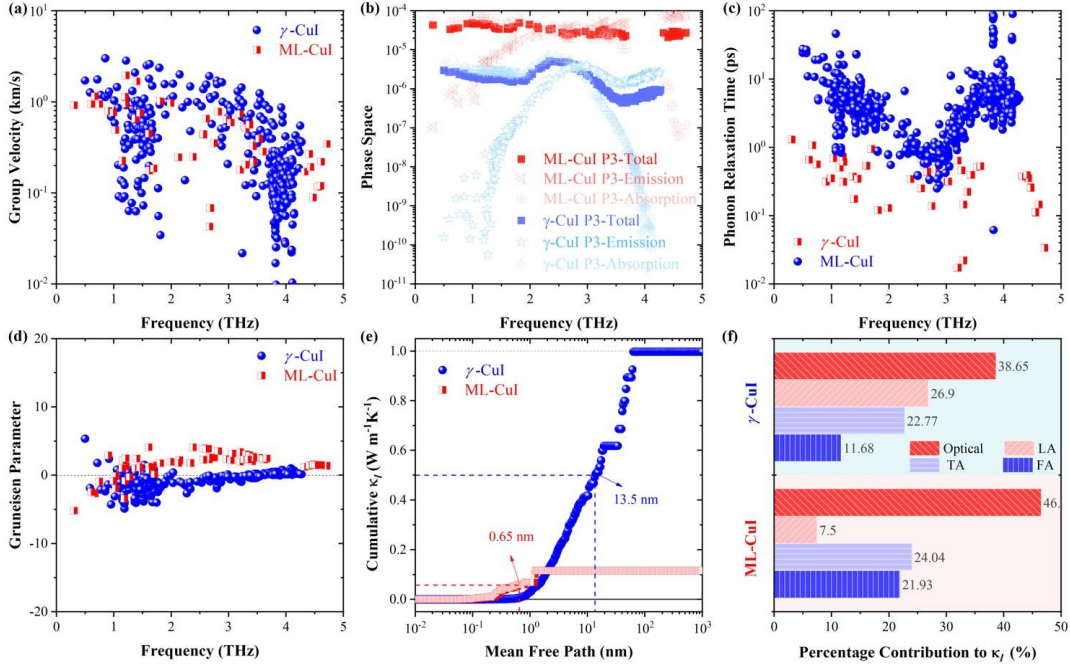

Figure. 3. The thermal transport properties of monolayer CuI and $\gamma$-CuI at 300 K. The modal analysis for (a) group velocity, (b) phase space, (c) phonon relaxation time, (d) Grüneisen parameter, and (e) size-dependent thermal conductivity. (f) The contribution from different branches of *out-of-plane z*-direction (ZA), traverse (TA), and longitudinal (LA) acoustic, and optical phonon branches.

The underlying mechanism for the ultralow thermal conductivity may lie in the heavy atoms of Cu and I, and also the buckled structure of monolayer CuI. Consequently, both the phonon group velocity [Fig. 3(a)] and relaxation time [Fig. 3(c)] are small and the scattering phase space [Fig.3(b)] is large. In addition, the phonon anharmonicity is found strong in monolayer CuI as quantified by the large Grüneisen parameters as shown in Fig. 3(d). More specifically, the total absolute value Grüneisen parameters of the monolayer CuI (1.6055) is significantly larger than that of $\gamma$-CuI (0.4828). In addition, the phonon mean free path (MFP) of monolayer CuI is lower than that of the $\gamma$-CuI as shown in Fig. 3(e). Thus, the ultralow thermal conductivity emerges in monolayer CuI, which would benefit the applications of power supply in wearable devices or Iot by promising excellent thermoelectric performance. Moreover, detailed analysis reveals that the large contribution of optical phonon modes in monolayer CuI [Fig. 3(f)] is due to the relatively small phase space at the corresponding frequencies, which is resulted from the bandgap among the optical phonon branches as revealed in Fig. 1(g).

### 2.3 The direct wide bandgap and visible transparency

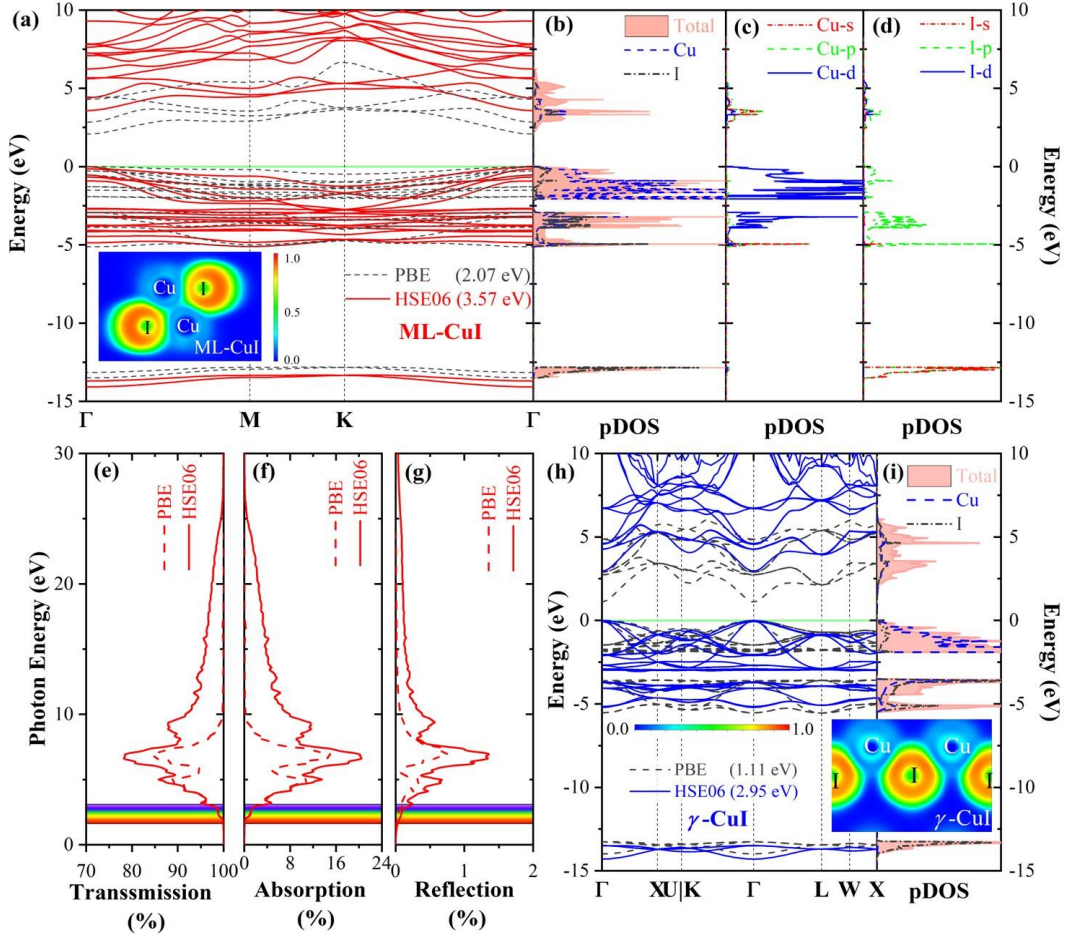

Figure 4. The electronic structures of monolayer CuI in comparison to $\gamma$-CuI. The electronic band structures are calculated by PBE and HSE06 for (a) monolayer CuI and (h) $\gamma$-CuI, where the electron localization functions (ELF) are presented as insets. The orbital projected density of states (pDOS) calculated by PBE of (b-d) monolayer CuI and (i) $\gamma$-CuI. The (e) transmission, (f) absorption, and (g) reflection coefficient of monolayer CuI.

With the verified stable structure of monolayer CuI, the electronic structures are further investigated as shown in Fig. 4(a). Using the PBE functional, a direct bandgap is predicted in monolayer CuI, and both the conduction band minimum (CBM) and valence band maximum (VBM) locate at the Gamma point. The bandgap is predicted as 2.07 eV by PBE. It is widely recognized that the bandgap is usually underestimated by PBE[26]. To accurately predict the bandgap of monolayer CuI, we utilize the HSE06 and obtain the bandgap to be 3.57 eV. The bandgap calculated by HSE06 keeps being direct similar to BPE, which promises the potential applications of monolayer CuI in electronics and optoelectronics since phonon is not required for electron transition. Moreover, the bandgap of monolayer CuI (3.57 eV) is larger than that of bulk $\gamma$-CuI (2.95 eV) as shown in Fig. 4 (h), which is in good agreement with 3.17 eV reported by Mustonen, K. *et al*[21], making monolayer CuI a promising direct wide bandgap semiconductor. In addition, the VBM is found to be mainly contributed by the Cu-*d* orbital as indicated by the pDOS in Fig. 4(b-d).

As shown in Fig. 4(e-g), in the energy domain of 0-7 eV, the absorption [Fig. 4(f)] and refraction [Fig. 4(g)] coefficients of monolayer CuI increase continuously, which means that monolayer CuI has an enhanced ability to absorb and refract light in this region. Correspondingly, the photon transmission [Fig. 4(e)] capability of monolayer CuI becomes weaker as the transmission coefficient decreases. When the photon energy is larger than 7 eV, the absorption and refraction coefficients of CuI start to be weakened significantly, and finally reach a plateau at the energy threshold of 8 eV. It is worth noting that, compared with the absorption and transmission capabilities of phonons, the reflection efficiency of monolayer CuI for photons is low, up to no more than 2%. For photon absorption, the working region of monolayer CuI is in the energy region of 5.0-7.5 eV, while the photon energy of visible light is between 1.62 and 3.11 eV. Obviously, the main absorption light of CuI is ultraviolet light, up to 20%. In short, the ultra-low thermal conductivity and ultra-high optical transmission enable monolayer CuI to develop great potential applications in the field of photothermal materials[27].

## 3. Conclusion

In summary, from the *state-of-art* first-principles, by examining the stability of monolayer CuI in different patterns of 2D structures, we have predicted the stable structure of monolayer

CuI as the counterpart of bulk $\gamma$-CuI, which possesses two buckled sublayers. We presented the multifunctional superiority of monolayer CuI by performing systematical study on the electronic, optical, and thermal transport properties. A direct bandgap is found in monolayer CuI (3.57 eV), which is larger than the bulk $\gamma$-CuI (2.95 eV), leading to the better optical performance. Moreover, the predicted ultralow thermal conductivity is close to that of air for monolayer CuI (0.116 W $m^{-1}K^{-1}$), which is only ~1/10 of its bulk counterpart $\gamma$-CuI (0.997 W $m^{-1}K^{-1}$) and much lower than the common semiconductors. The direct wide bandgap and the ultralow thermal conductivity of monolayer CuI would promise its potential applications in transparent and wearable electronics.

## 4. Computational methodology

All the first-principles calculations are implemented by the Vienna ab initio simulation package (VASP)[28] within the framework of density functional theory (DFT). The interaction between valance electrons and ion cores is performed by Projector augmented wave (PAW)[29] method. The Perdew-Burke-Ernzerhof (PBE) functional was chosen as the exchange-correlation functional for geometric structure optimization and thermal transport properties. Furthermore, the hybridized HSE06 functional was used to obtain precise electrical and optical properties. To prevent the periodic mirroring interaction along out-of-plane direction, 16 Å is used as a vacuum layer. The 920 eV is selected as the kinetic energy cutoff to expand the wave functions for plane wave basis with a Gamma centered grid[30] k-meshes of 8×8×1 to sample the irreducible Brillouin zone (IBZ). For geometric optimization, the energy convergence threshold of $10^{-8}$ eV implemented until the Hellmann-Feynman force convergence accuracy of $-5\times10^{-7}$ eV/Å. For $2^{nd}$- and $3^{rd}$-order interatomic force constants (IFCs) calculations, 6×6×1 and 2×2×2 supercell containing 144 and 64 atoms are used for monolayer CuI and $\gamma$-CuI, respectively.

The linear optical properties of monolayer CuI are investigated based on the precise electronic structures obtained from HSE06, and the absorption coefficient $\alpha(\omega)$, refractive coefficient $R(\omega)$ and reflection coefficient $n(\omega)$ of CuI can be obtained [Fig. 2 (e-g)] as[31–33]:

$$\alpha(\omega) = \frac{2\omega k(\omega)}{c} = \frac{\varepsilon_2(\omega)\omega}{n(\omega)c} \tag{1}$$

$$R(\omega) = \frac{[n(\omega)-1]^2 + k(\omega)^2}{[n(\omega)+1]^2 + k(\omega)^2} \tag{2}$$

$$n(\omega) = \frac{1}{\sqrt{2}}\left[\sqrt{\varepsilon_1(\omega)^2 + \varepsilon_2(\omega)^2} + \varepsilon_1(\omega)\right]^{\frac{1}{2}} \tag{3}$$

where $\varepsilon_1(\omega)$ and $\varepsilon_2(\omega)$ are the real and imaginary parts of the dielectric function, respectively. The term $k(\omega)$ is the extinction coefficient, which can be derived by[31,32]:.

$$k(\omega) = \frac{1}{\sqrt{2}}\left[\sqrt{\varepsilon_1(\omega)^2 + \varepsilon_2(\omega)^2} - \varepsilon_1(\omega)\right]^{\frac{1}{2}} \tag{4}$$

The real $\varepsilon_1(\omega)$ can be obtained by the Kramers–Kronig relation[31]:

$$\varepsilon_1(\omega) = 1 + \frac{2}{\pi}P\int_0^\infty \frac{\varepsilon_2(\alpha)\alpha d\alpha}{\alpha^2 - \omega^2} \tag{5}$$

and imaginary parts $\varepsilon_2(\omega)$ can be derived by[34] :

$$\varepsilon_2(\omega) = \frac{4\pi^2 e^2}{m^2\pi^2} \times \sum_{C,V}\left|M_{C,V}\right|^2 \delta(E_C + E_V - h\omega) \tag{6}$$

where the terms $e$, $m$, and $\omega$ are electron charge, effective mass, and angular frequency, respectively. The terms $C$ and $V$ represent the conduction band and valence band, respectively. The term $M$ represents the momentum transfer matrix. The term $E$ is the electron energy level, and the delta function $\delta$ can enhance the energy conversion of electrons during the transition from band to band.

The thermal conductivity is obtained as implemented in the ShengBTE package[35,36] by summing all the phonon mode ($\lambda$) contributions:

$$\kappa = \kappa_{\alpha\alpha} = \frac{1}{V}\sum_\lambda C_\lambda v_{\lambda\alpha}^2 \tau_{\lambda\alpha} \tag{7}$$

where $V$ is the lattice volume associated with the effective thickness, and 7.81 Å is chosen as the effective thickness for monolayer CuI. In addition, $C_\lambda$, $v_{\lambda\alpha}$, and $\tau_{\lambda\alpha}$ are the specific heat capacity, group velocity and relaxation time. Moreover, they can be further obtained through lattice dynamics. The phonon scattering rate is inversely proportional to the relaxation time, *i.e.*, $\Gamma = 1/\tau$. The scattering rate is obtained by summing the different scattering terms based on Matthiessen's rule[37]:

$$\frac{1}{\tau(\vec{q},p)} = \frac{1}{\tau^{anh}(\vec{q},p)} + \frac{1}{\tau^{iso}(\vec{q},p)} + \frac{1}{\tau^{B}(\vec{q},p)} \tag{8}$$

where $1/\tau^{anh}$, $1/\tau^{iso}$ and $1/\tau^{B}$ is the intrinsic three-phonon scattering rate, the scattering rate due to isotope impurities, and boundary scattering rate due to the finite length of sample. More importantly, the three-phonon scattering rate $\Gamma^{\pm}_{\lambda\lambda'\lambda''}$ can be further obtained by[5,7–9]:

$$\Gamma^{+}_{\lambda\lambda'\lambda''} = \frac{\hbar\pi}{4} \frac{f_0' - f_0''}{\omega_\lambda \omega_{\lambda'} \omega_{\lambda''}} \left| V^{+}_{\lambda\lambda'\lambda''} \right|^2 \delta\left( \omega_{\lambda''} + \omega_{\lambda'} - \omega_\lambda \right) \tag{9}$$

$$\Gamma^{-}_{\lambda\lambda'\lambda''} = \frac{\hbar\pi}{4} \frac{f_0' + f_0'' + 1}{\omega_\lambda \omega_{\lambda'} \omega_{\lambda''}} \left| V^{-}_{\lambda\lambda'\lambda''} \right|^2 \delta\left( \omega_{\lambda''} - \omega_{\lambda'} - \omega_\lambda \right) \tag{10}$$

where the scattering rate $\Gamma^{+}_{\lambda\lambda'\lambda''}$ and $\Gamma^{-}_{\lambda\lambda'\lambda''}$ are determined by the absorption emission process, respectively. In addition, $\hbar$, $\omega$, $V^{\pm}_{\lambda\lambda'\lambda''}$, $\delta$, and $f_0$ are Planck's constant, phonon frequency, the scattering matrix elements, the Dirac delta distribution function and the Bose-Einstein function, respectively.

## Acknowledgments

This work is supported by the National Natural Science Foundation of China (Grant No. 52006057), the Fundamental Research Funds for the Central Universities (Grant Nos. 531119200237, and 541109010001), and the State Key Laboratory of Advanced Design and Manufacturing for Vehicle Body at Hunan University (Grant No. 52175011). The numerical calculations in this paper have been done on the supercomputing system of the National Supercomputing Center in Changsha and Zhengzhou.

**Note**: During the preparation of the manuscript, we noticed that the monolayer CuI was synthesized very recently (*Adv. Mater.* 34, 2106922 (2022)), which has the same structure as predicted in this work.